# Considerations and recommendations from the ISMRM Diffusion Study Group for preclinical diffusion MRI: Part 2 — *Ex vivo* imaging: added value and acquisition

Kurt G Schilling[1,2,#], Francesco Grussu[3,4], Andrada Ianus[5], Brian Hansen[6], Amy FD Howard[7,8], Rachel L C Barrett[9,10], Manisha Aggarwal[11], Stijn Michielse[12], Fatima Nasrallah[13], Warda Syeda[14], Nian Wang[15,16], Jelle Veraart[17], Alard Roebroeck[18], Andrew F Bagdasarian[19,20], Cornelius Eichner[21], Farshid Sepehrband[22], Jan Zimmermann[23], Lucas Soustelle[24], Christien Bowman[25,26], Benjamin C Tendler[27], Andreea Hertanu[28], Ben Jeurissen[29,30], Marleen Verhoye[25,26], Lucio Frydman[31], Yohan van de Looij[32], David Hike[19,20], Jeff F Dunn[33,34,35], Karla Miller[8], Bennett A Landman[36], Noam Shemesh[5], Adam Anderson[37,2], Emilie McKinnon[38], Shawna Farquharson[39], Flavio Dell' Acqua[40], Carlo Pierpaoli[41], Ivana Drobnjak[42], Alexander Leemans[43], Kevin D Harkins[1,2,44], Maxime Descoteaux[45,46], Duan Xu[47], Hao Huang[48,49], Mathieu D Santin[50,51], Samuel C. Grant[19,20], Andre Obenaus[52,53], Gene S Kim[54], Dan Wu[55], Denis Le Bihan[56,57], Stephen J Blackband[58,59,60], Luisa Ciobanu[61], Els Fieremans[62], Ruiliang Bai[63,64], Trygve B Leergaard[65], Jiangyang Zhang[66], Tim B Dyrby[67,68], G Allan Johnson[69,70], Julien Cohen-Adad[71,72,73], Matthew D Budde[74,75], Ileana O Jelescu[28,76,#]

[#]Corresponding authors — kurt.g.schilling.1@vumc.org, ileana.jelescu@chuv.ch

[1]Radiology and Radiological Sciences, Vanderbilt University Medical Center, Nashville, TN, USA, [2]Vanderbilt University Institute of Imaging Science, Vanderbilt University, Nashville, TN, [3]Radiomics Group, Vall d'Hebron Institute of Oncology, Vall d'Hebron Barcelona Hospital Campus, Barcelona, Spain, [4]Queen Square MS Centre, Queen Square Institute of Neurology, Faculty of Brain Sciences, University College London, London, UK, [5]Champalimaud Research, Champalimaud Foundation, Lisbon, Portugal, [6]Center of Functionally Integrative Neuroscience, Aarhus University, Aarhus, Denmark, [7]Department of Bioengineering, Imperial College London, London, UK, [8]FMRIB Centre, Wellcome Centre for Integrative Neuroimaging, Nuffield Department of Clinical Neurosciences, University of Oxford, Oxford, United Kingdom, [9]Department of Neuroimaging, Institute of Psychiatry, Psychology and Neuroscience, King's College London, London, UK, [10]NatBrainLab, Department of Forensics and Neurodevelopmental Sciences, Institute of Psychiatry, Psychology and Neuroscience, King's College London, London, UK, [11]Russell H. Morgan Department of Radiology and Radiological Science, Johns Hopkins University School of Medicine, Baltimore, MD, USA, [12]Department of Neurosurgery, School for Mental Health and Neuroscience (MHeNS), Maastricht University Medical Center, Maastricht, The Netherlands, [13]The Queensland Brain Institute, The University of Queensland, Queensland, Australia, [14]Melbourne Neuropsychiatry Centre, The University of Melbourne, Parkville, Victoria, Australia, [15]Department of Radiology and Imaging Sciences, Indiana University, IN, USA, [16]Stark Neurosciences Research Institute, Indiana University School of Medicine, IN, USA, [17]Center for Biomedical Imaging, NYU Grossman School of Medicine, New York, NY, USA, [18]Faculty of psychology and Neuroscience, Maastricht University, Maastricht, Netherlands, [19]Department of Chemical & Biomedical Engineering, FAMU-FSU College of Engineering, Florida State University, Tallahassee, FL, USA, [20]Center for Interdisciplinary Magnetic Resonance, National HIgh Magnetic Field

Laboratory, Tallahassee, FL, USA, [21]Department of Neuropsychology, Max Planck Institute for Human Cognitive and Brain Sciences, Leipzig, Germany, [22]USC Stevens Neuroimaging and Informatics Institute, Keck School of Medicine of USC, University of Southern California, Los Angeles, CA, USA, [23]Department of Neuroscience, Center for Magnetic Resonance Research, University of Minnesota, MN, USA, [24]Aix Marseille Univ, CNRS, CRMBM, Marseille, France, [25]Bio-Imaging Lab, Faculty of Pharmaceutical, Biomedical and Veterinary Sciences, University of Antwerp, Antwerp, Belgium, [26]μNEURO Research Centre of Excellence, University of Antwerp, Antwerp, Belgium, [27]Wellcome Centre for Integrative Neuroimaging, FMRIB, Nuffield Department of Clinical Neurosciences, University of Oxford, United Kingdom, [28]Department of Radiology, Lausanne University Hospital and University of Lausanne, Lausanne, Switzerland, [29]imec Vision Lab, Dept. of Physics, University of Antwerp, Belgium, [30]Lab for Equilibrium Investigations and Aerospace, Dept. of Physics, University of Antwerp, Belgium, [31]Department of Chemical and Biological Physics, Weizmann Institute of Science, Rehovot, Israel, [32]Division of Child Development & Growth, Department of Pediatrics, Gynaecology & Obstetrics, School of Medicine, Université de Genève, Genève, Switzerland, [33]Department of Radiology, Cumming School of Medicine, University of Calgary, Calgary, Alberta, Canada, [34]Hotchkiss Brain Institute, Cumming School of Medicine, University of Calgary, Calgary, Alberta, Canada, [35]Alberta Children's Hospital Research Institute, Cumming School of Medicine, University of Calgary, Calgary, Alberta, Canada, [36]Department of Electrical and Computer Engineering, Vanderbilt University,, [37]Department of Radiology and Radiological Sciences, Vanderbilt University Medical Center, Nashville, TN, USA, [38]Medical University of South Carolina, Charleston, SC, USA, [39]National Imaging Facility, The University of Queensland, Brisbane, Australia, [40]Department of Forensic and Neurodevelopmental Sciences, King's College London, London, UK, [41]Laboratory on Quantitative Medical imaging, NIBIB, National Institutes of Health, Bethesda, MD, USA, [42]Department of Computer Science, University College London, London, UK, [43]PROVIDI Lab, Image Sciences Institute, University Medical Center Utrecht, The Netherlands, [44]Biomedical Engineering, Vanderbilt University, Nashville, TN, [45]Sherbrooke Connectivity Imaing Lab (SCIL), Computer Science department, Université de Sherbrooke, [46]Imeka Solutions, [47]Department of Radiology and Biomedical Imaging, University of California San Francisco, CA, USA, [48]Department of Radiology, Perelman School of Medicine, University of Pennsylvania, Philadelphia, PA, USA, [49]Department of Radiology, Children's Hospital of Philadelphia, Philadelphia, PA, USA, [50]Centre for NeuroImaging Research (CENIR), Inserm U 1127, CNRS UMR 7225, Sorbonne Université, Paris, France, [51]Paris Brain Institute, Paris, France, [52]Department of Pediatrics, University of California Irvine, Irvine CA USA, [53]Preclinical and Translational Imaging Center, University of California Irvine, Irvine CA USA, [54]Department of Radiology, Weill Cornell Medical College, New York, NY, USA, [55]Key Laboratory for Biomedical Engineering of Ministry of Education, College of Biomedical Engineering & Instrument Science, Zhejiang University, Hangzhou, China, [56]CEA, DRF, JOLIOT, NeuroSpin, Gif-sur-Yvette, France, [57]Université Paris-Saclay, Gif-sur-Yvette, France, [58]Department of Neuroscience, University of Florida, Gainesville, FL, United States, [59]McKnight Brain Institute, University of Florida, Gainesville, FL, United States, [60]National High Magnetic Field Laboratory, Tallahassee, FL, United States, [61]NeuroSpin, UMR CEA/CNRS 9027, Paris-Saclay University, Gif-sur-Yvette, France, [62]Department of Radiology, New York University Grossman School of Medicine, New York, NY, USA, [63]Interdisciplinary Institute of Neuroscience and Technology, School of Medicine, Zhejiang University, Hangzhou, China, [64]Frontier Center of Brain Science and Brain-machine Integration, Zhejiang University, [65]Department of Molecular Biology, Institute of Basic Medical Sciences, University of Oslo, Norway, [66]Department of Radiology, New York University School of Medicine, NY, NY, USA, [67]Danish Research Centre for Magnetic Resonance, Centre for Functional and Diagnostic Imaging and Research, Copenhagen University Hospital Amager & Hvidovre, Hvidovre, Denmark, [68]Department of Applied Mathematics and Computer Science, Technical University of Denmark, Kongens Lyngby, Denmark, [69]Duke Center for In Vivo Microscopy, Department of Radiology, Duke University, Durham, North Carolina, [70]Department of Biomedical Engineering, Duke University, Durham, North Carolina, [71]NeuroPoly Lab, Institute of Biomedical Engineering, Polytechnique Montreal, Montreal, QC, Canada, [72]Functional Neuroimaging Unit, CRIUGM, University of Montreal, Montreal, QC, Canada, [73]Mila - Quebec AI Institute, Montreal, QC, Canada, [74]Department of Neurosurgery, Medical College of Wisconsin, Milwaukee, Wisconsin, [75]Clement J Zablocki VA Medical Center, Milwaukee, Wisconsin, [76]CIBM Center for Biomedical Imaging, Ecole Polytechnique Fédérale de Lausanne, Lausanne, Switzerland

# Abstract

The value of preclinical diffusion MRI (dMRI) is substantial. While dMRI enables *in vivo* non-invasive characterization of tissue, *ex vivo* dMRI is increasingly being used to probe tissue microstructure and brain connectivity. *Ex vivo* dMRI has several experimental advantages including higher signal-to-noise ratio (SNR) and spatial resolution compared to *in vivo* studies, and enabling more advanced diffusion contrasts for improved microstructure and connectivity characterization. Another major advantage of *ex vivo* dMRI is the direct comparison with histological data, as a crucial methodological validation. However, there are a number of considerations that must be made when performing *ex vivo* experiments. The steps from tissue preparation, image acquisition and processing, and interpretation of results are complex, with many decisions that not only differ dramatically from *in vivo* imaging of small animals, but ultimately affect what questions can be answered using the data. This work represents "Part 2" of a 3-part series of recommendations and considerations for preclinical dMRI. We describe best practices for dMRI of *ex vivo* tissue, with a focus on the value that ex vivo imaging adds to the field of dMRI and considerations in ex vivo image acquisition. We first give general considerations and foundational knowledge that must be considered when designing experiments. We briefly describe differences in specimens and models and discuss why some may be more or less appropriate for different studies. We then give guidelines for *ex vivo* protocols, including tissue fixation, sample preparation, and MR scanning. In each section, we attempt to provide guidelines and recommendations, but also highlight areas for which no guidelines exist (and why), and where future work should lie. An overarching goal herein is to enhance the rigor and reproducibility of ex vivo dMRI acquisitions and analyses, and thereby advance biomedical knowledge.

# 1 Introduction

Diffusion magnetic resonance imaging (dMRI) is a medical imaging technique that utilizes the diffusion of water molecules to generate image contrast, enabling the non-invasive mapping of the diffusion process in biological tissues. These diffusion patterns can be used to infer and generate maps of tissue microstructure, or can be applied to map tissue orientation to study the structural connections of the brain in a process called fiber tractography. They have found applications widely used in neuroscience, neuroanatomy, and neurology, but also outside the brain with musculoskeletal, whole-body, and organ-specific applications in normal and pathological conditions.

The use of animal models and *ex vivo* tissue is essential to the field of diffusion MRI. In this work, **we define *ex vivo* as covering any fresh excised tissue, perfused living tissue, or fixed tissue.** While dMRI enables non-invasive characterization of tissue *in vivo*, *ex vivo* dMRI acquisitions have several experimental advantages, including longer scanning times and absence of motion.

However, there are a number of considerations that must be made when performing *ex vivo* experiments.The steps from tissue preparation, image acquisition and processing, and interpretation of results are complex, with many decisions that not only differ dramatically from *in vivo* imaging of small animals, but ultimately affect what questions can be answered using the data. This work represents “Part 2” of a three-part series of recommendations and considerations for preclinical dMRI. Part 1 covers *in vivo* small animal imaging. Here Part 2 presents general considerations and best practices for preclinical dMRI acquisition of *ex vivo* tissue. Finally, Part 3 covers ex vivo data processing, tractography, and comparisons with histology. **This work does not serve as a “consensus” on any specific topic, but rather as a snapshot of “best practices” or “guidance”** from the preclinical dMRI community as represented by the authors. We envision this work to be useful to imaging centers using small animal scanners for research, sites that may not have personnel with expert knowledge in diffusion, pharmaceutical or industry employees, or new trainees in the field of dMRI. The resources provided herein may act as a starting point when reading the literature, and understanding the decisions and processes for studying model systems with dMRI.

We first describe the value that *ex vivo* imaging adds to the field of dMRI, followed by general considerations and foundational knowledge that must be considered when designing experiments. We briefly describe differences in specimens and models and discuss why some may be more or less appropriate for different studies. We then provide guidance for *ex vivo* acquisition protocols, including decisions on hardware, sample preparation, and imaging sequences. In each section, we attempt to also highlight areas for which no guidelines exist (and why), and where future work should lie. With this, we hope to enhance the rigor and reproducibility of ex vivo dMRI acquisitions and analyses, and thereby advance biomedical knowledge.

# 2 Added Value

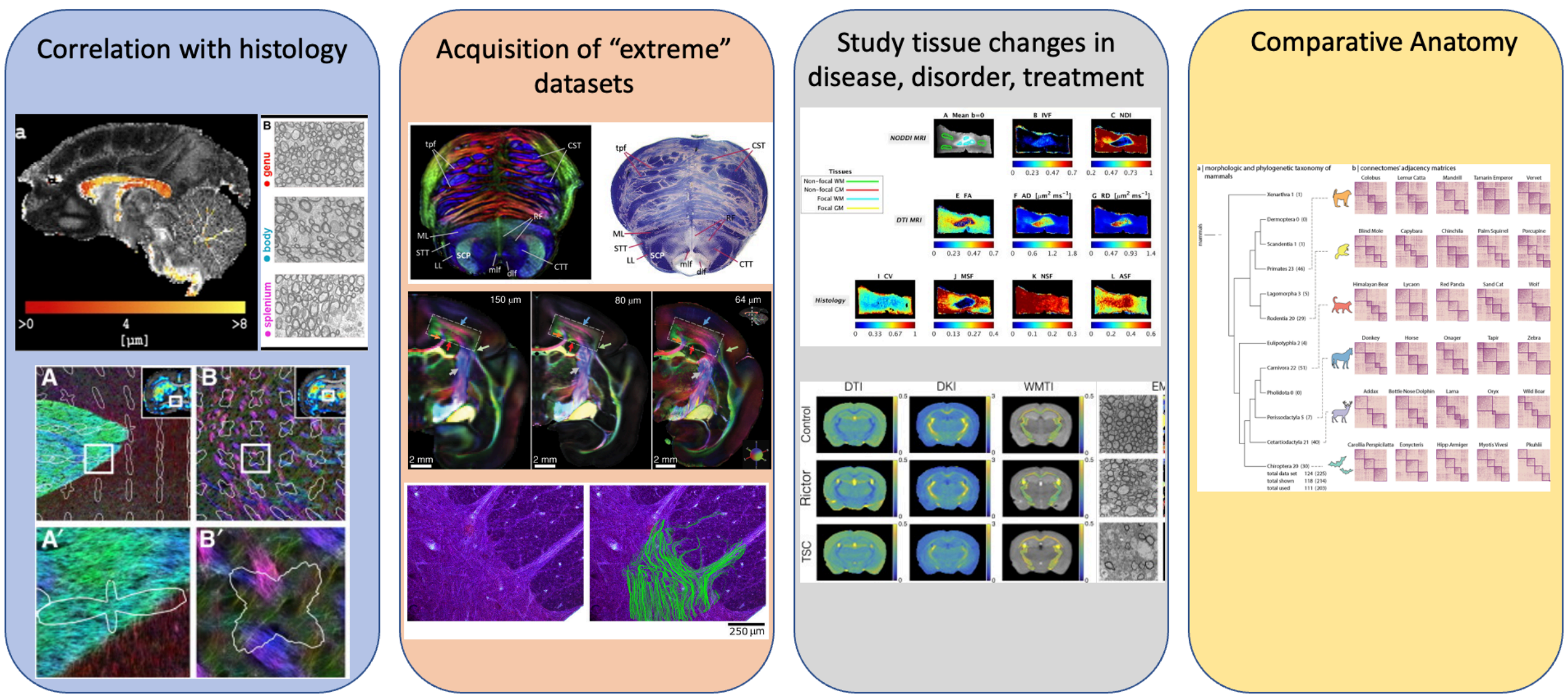

**Figure 1**. Four areas in which preclinical brain imaging adds value to the field of dMRI. It enables: (i) correlation with histology on the same subject/sample, (ii) the acquisition of richer datasets than on clinical systems thanks to more advanced hardware and longer scan times available, (iii) the study of tissue changes with disease and treatment in a more controlled setting, and (iv) comparative anatomy between species. Figures reused and adapted from (left to right): (i) [1–3] (ii) [4–6] (iii) [7,8] (iv) [9].

*Ex vivo* samples add substantial value to dMRI by acting as supplements, substitutes, translations, and/or validation mechanisms for *in vivo* human studies. Building on the four areas identified in Part 1 in which preclinical MRI adds value to the field of diffusion MRI, here we focus on aspects specific to *ex vivo* experiments (**Figure 1**).

**<u>First, *ex vivo* dMRI allows correlations with histological and other imaging measures</u>** in a more direct way than *in vivo* dMRI, since the tissue has undergone the similar changes related to fixation and other chemical treatments. Through co-registration of microscopy images to MRI images of the same specimens, direct comparisons and correlation of dMRI measures to different quantitative microscopic parameters can be achieved, thus elucidating how different microscopic tissue features influence dMRI contrast. Such validation studies in turn improve our ability to interpret *in vivo* dMRI for both preclinical and clinical studies.

**<u>Second, *ex vivo* imaging allows acquisition of “extreme” datasets</u>** not possible with *in vivo* preclinical or clinical imaging, pushing the boundaries of acquisition and analysis to answer questions about what dMRI is capable of measuring in principle. By “extreme” datasets, we target in particular: advanced diffusion encoding, higher *b*-values, shorter diffusion times, or simply very comprehensive *q-t* coverage, all with high signal-to-noise ratio (SNR) and/or high spatial resolution that requires very long scan times. Preclinical MRI systems indeed offer four distinct advantages for imaging.

First, the appeal of **higher fields** is driven by benefits of potentially higher SNR due to increased net magnetization. Second, ex vivo imaging allows more flexibility in **RF coils**, which can be in close proximity to the sample being imaged, further increasing SNR. Third, **stronger and faster switching gradients** are also an immense asset for dMRI experiments [10], enabling a higher SNR via reduced echo time. Stronger gradients facilitate the exploration of broader ranges of diffusion sensitization and diffusion time (i.e. "*q-t* space") while keeping an acceptable SNR level. In particular, unique insights into tissue microstructure have been brought by exploring a range of *q-t* space only feasible (at the time) on preclinical systems [11–15], including very short diffusion times, very strong diffusion weightings [11–14; 2,16,17], or more complex diffusion encoding schemes [18–20]. Finally, the greatest asset enabled by *ex vivo* imaging is a lengthy **scan time**, ranging from several hours to successful scans of >10 days in duration, with the added benefit of no physiological motion. Long scanning further facilitates acquisitions across *q-t* ranges, with substantially increased SNR and/or increased spatial resolution.

**Third, the use of animal models allows us to study the sensitivity of dMRI to tissue changes in diseases, disorders and treatments in a controlled way**. Animal models are critical to biomedical research as they may be biologically similar to humans, susceptible to many of the same health problems, and can be studied throughout their whole life span and across generations. E*x vivo* imaging has the added benefit of bridging the gap between *in vivo* imaging and histology (Added Value #1) of such models, and investigation of new imaging sequences (Added Value #2) that together spur development and validation of new biomarkers for diagnosis and treatment.

**Fourth, the use of *ex vivo* systems enables comparative anatomy**. A key challenge in comparative neuroanatomy is to identify homologous structures and structural boundaries across species. Moreover, the brain undergoes substantial changes through development and aging which hampers comparison of data from different timepoints. High-resolution *ex vivo* structural MRI and dMRI data have provided versatile reference data for creating anatomical atlases for fly, mouse, rat, primate, and even bat brains [21–28] allowing detection of detailed anatomical systems corresponding to those identified using histological criteria [29,30]. Comparative neuroanatomical efforts, then, may rely on histological comparisons across species supplemented by high quality imaging acquisitions as a marker of 'virtual histology' and 'virtual brain dissection'.

# 3 *Ex vivo*: Translation and validation considerations

Similar to *in vivo* studies of small animals, scanning *ex vivo* tissue presents both opportunities and challenges when translating and validating experimental findings [31]. This section introduces experimental and biological aspects that must be considered when designing and interpreting *ex vivo* studies. All aspects are covered in more detail in their corresponding sub-sections in *Acquisition* (Section 4).

**Anatomical considerations**

The basic constituents of the brain and other organs are largely preserved across mammalian species, providing the basis for translational MRI studies. Moreover, the constituents and organization of *ex vivo* tissue are largely similar to their *in vivo* counterparts, assuming appropriate perfusion fixation (to preserve 'dead' tissue) or artificial perfusion (to preserve 'viable' *ex vivo* tissue) [32,33]. For example, in the central nervous system, the fundamental structure of a long axon wrapped with a myelin sheath is preserved with chemical fixation *ex vivo*. Noticeable exceptions to similarity are possible changes in size, volume fractions of tissue compartments, membrane permeability and/or diffusivity drop beyond those that are expected from temperature changes (see *Considerations of microstructure and the diffusion process*, below) [34,35].

**Considerations in disease/disorder/model**

As in small animal studies, a challenge when using *ex vivo* data to study the sensitivity of dMRI to detect changes in disease or treatments is confirming the translational value to human studies. Despite this, *ex vivo* imaging of models of stroke, demyelination, traumatic brain injury, spinal cord injury, and tumor models have proven useful to investigate altered microstructure or connectivity in diseased states, and facilitate subsequent histological validation [36]. Clearly, the advantage of small animal imaging for longitudinal studies of development or disease progression in single subjects stops at the *ex vivo* scan, yet the benefits of the high resolution and high SNR scan enable detailed investigations of the *ex vivo* tissue at one specific time point.

**Considerations of microstructure and the diffusion process**

The translation of *ex vivo* measurements to *in vivo* should be interpreted with caution, as both the tissue microstructure and intrinsic diffusivities may be different (**Figure 2**). To prevent postmortem degradation and to limit uncontrolled changes in tissue microstructure, tissues should be chemically fixed. Nevertheless, even using appropriate **chemical fixation**, the tissue inevitably undergoes changes: so far intra-/extracellular space volume fraction shifts, cell membrane permeability, relaxation rates and diffusion coefficients have all been shown to change [34,35,37–39]. The degree to which fixation affects microstructure features is however variable across the literature - see Section 4.3 for details. Ex vivo dMRI on fixed tissue at room temperature also distinguishes itself through **lower diffusivity by a factor of 2-5** in the tissue, which cannot be explained by the difference in temperature alone, but is also due to fixation and/or other post-mortem tissue changes [40]. These effects should all be considered when designing the scanning protocol, and should be considered also when interpreting results in terms of diffusion distance and probed spatial scales (Section 4.5.3). Lastly, strategies for mitigating and/or correcting **temporal drift** are described throughout, and may involve sample preparation (Section 4.4), controlling temperature during scanning (Section 4.5.4) and within preprocessing pipelines (Part 3; Section 2).

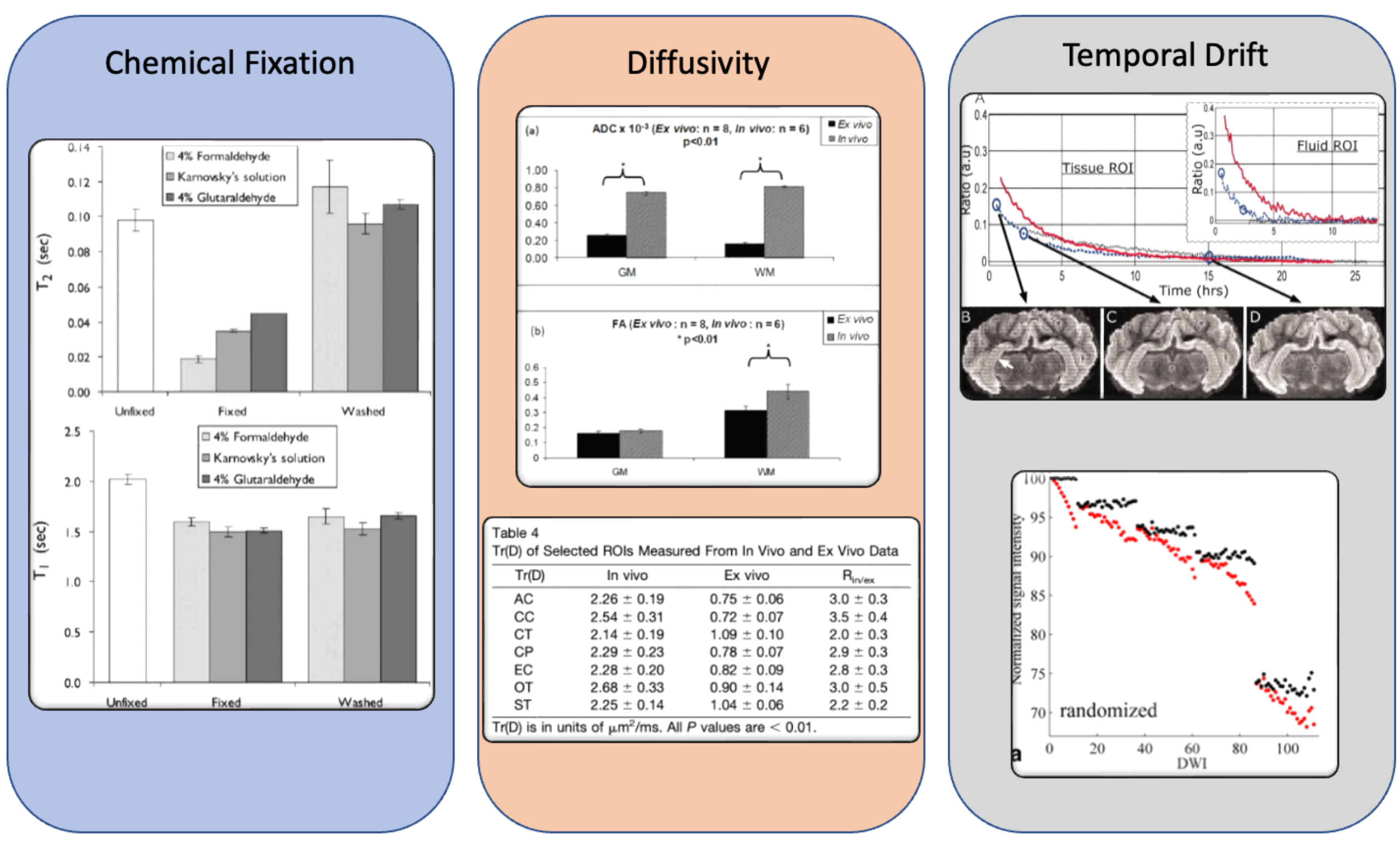

Table 4
Tr(D) of Selected ROIs Measured From In Vivo and Ex Vivo Data

| Tr(D) | In vivo | Ex vivo | $R_{in/ex}$ |
|---|---|---|---|
| AC | 2.26 ± 0.19 | 0.75 ± 0.06 | 3.0 ± 0.3 |
| CC | 2.54 ± 0.31 | 0.72 ± 0.07 | 3.5 ± 0.4 |
| CT | 2.14 ± 0.19 | 1.09 ± 0.10 | 2.0 ± 0.3 |
| CP | 2.29 ± 0.23 | 0.78 ± 0.07 | 2.9 ± 0.3 |
| EC | 2.28 ± 0.20 | 0.82 ± 0.09 | 2.8 ± 0.3 |
| OT | 2.68 ± 0.33 | 0.90 ± 0.14 | 3.0 ± 0.5 |
| ST | 2.25 ± 0.14 | 1.04 ± 0.06 | 2.2 ± 0.2 |

Tr(D) is in units of $\mu m^2/ms$. All *P* values are $< 0.01$.

**Figure 2** Considerations in the diffusion process. When performing studies on *ex vivo* tissue, one must consider effects of (i) chemical fixation (changes in geometry, volume fractions, relaxation rates, permeability, diffusion coefficients), (ii) changes in diffusivity (which can be ~2-5 times reduced from *in vivo* depending on experimental conditions), and (iii) temporal instabilities over long scan times (causing temporal drift or image artifacts). Figures reused and adapted from (i) [35] (ii) [41,42] (iii) [43,44]

These effects need to be considered when interpreting *in vivo* dMRI based on *ex vivo* validation studies. A diffusion biophysical model that is validated or performs well *ex vivo* may not be valid/validated *in vivo*, and vice-versa, due to modeling assumptions, and potential differences in diffusivities, compartmental volume fractions, and relaxation rates.

## 3.1 Species differences

As discussed in Part 1, dMRI has been utilized in a number of animal models, both *in vivo* and *ex vivo*. The most appropriate species to investigate is ultimately dependent upon the research question. Of course, in studies of human anatomy, the postmortem human brain can also be imaged and subsequently dissected, or sectioned for histological analysis. Additionally improved SNR can be achieved and traded for spatial resolution by dissecting and imaging just the fragment of interest from the whole sample, thus allowing the use of smaller-bore, higher-magnetic field systems, and smaller RF coils.

Below, we list some of the most common species (and a few avantgarde ones) studied with *ex vivo* dMRI and briefly describe advantages and disadvantages of each.

## 3.1.1 Murine models (mouse and rat)

Rats and mice have been and continue to be the long-standing preferred species for biomedical research models as they offer a low-cost option with outcome measures widely available and a substantial database of normative data, including behavioral, genomic, and medical imaging (**Figure 3**). In addition to biological advantages, the small physical size offers technical advantages, fitting in the typically smaller bores (and smaller coils) of magnets with larger field strengths. This is particularly advantageous for *ex vivo* imaging, where the small size may facilitate scanning with smaller bore ultra-high field scanners and smaller volume coils or cryogenic probes. *Ex vivo*, mouse and rat models have found most applications in exploration and development of advanced image acquisition and diffusion encoding, and validation of multi-compartment modeling facilitated through subsequent histology.

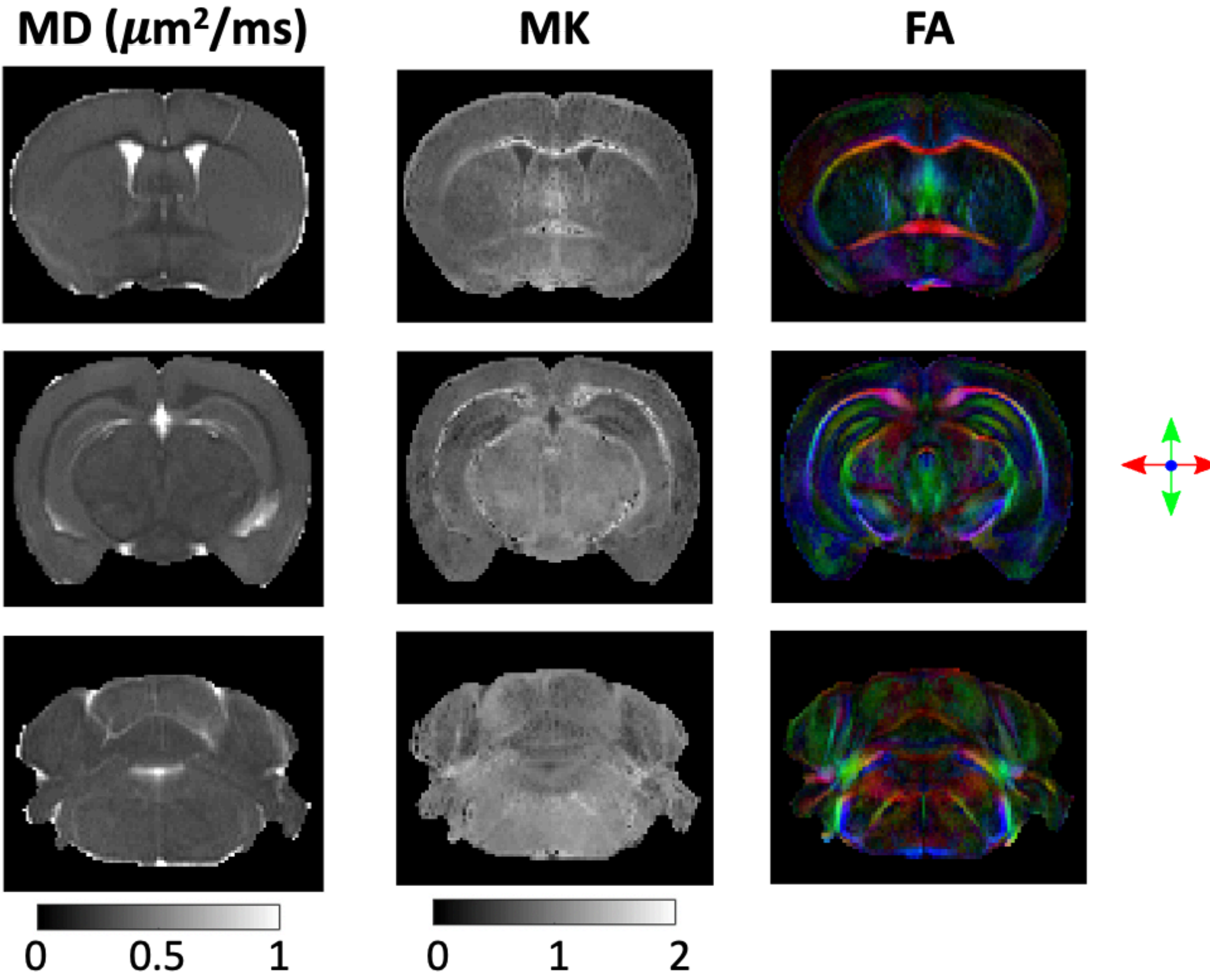

**Figure 3**. *Ex vivo* imaging of mouse models facilitates high resolution, high SNR, dense sampling of *q-t* space. Here, a fixed mouse brain was imaged on a 16.4T Bruker Aeon Ascend scanner equipped with a 10-mm birdcage coil and gradients capable of producing up to 3 T/m in all directions. Images show mean diffusivity (MD), Mean Kurtosis (MK), and directionally

encoded color (DEC) FA maps [45]. All animal studies were approved by the competent institutional and national authorities, and performed according to European Directive 2010/63. Images kindly provided by Andrada Ianus and Noam Shemesh.

### 3.1.2 Primate models

Popular non-human primates (NHPs) in dMRI literature include marmosets, squirrel monkeys, and macaques. With a number of white matter and gray matter regions with homologous counterparts in humans [46–49], NHPs are well suited for studies of cortical development, gyrification, and the structural and functional significance of specific white matter pathways. *Ex vivo* NHP imaging, in particular, is very common after injection of histological tracers, and constitutes a majority of diffusion tractography validation studies (**Figure 4**). This enables comparisons of tracers to exceptional quality diffusion datasets, in order to identify challenges and limitations in diffusion tractography. For example, controversies regarding the existence or nonexistence of a pathway, or the location of pathway terminations have been resolved or steered through primate models [50–53].

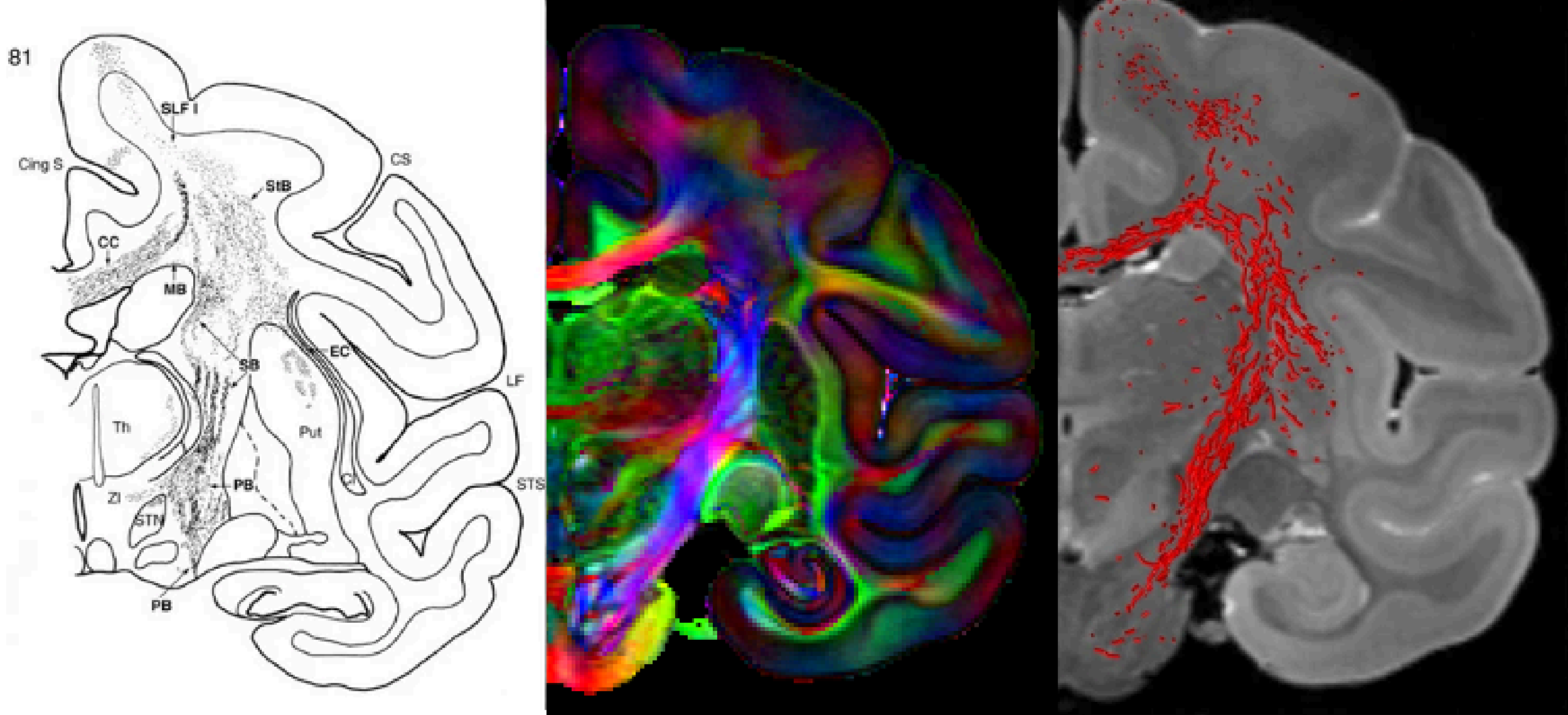

**Figure 4**. Primate models have been used to validate tractography estimates of structural connectivity. *Ex vivo* imaging offers the ability to investigate and compare the anatomical accuracy of high quality and high resolution dMRI datasets against histological tracers, the gold-standard for elucidating brain tractography. Figure adapted from [54] and [55], based on *ex vivo* macaque data acquired by [56] shows tracer trajectory (left), directionally encoded color map (middle) and tractography streamlines (right).

### 3.1.3 Human models

*Ex vivo* imaging of human brains is also possible (**Figure 5**). Of course, the greatest advantage is the immediate translatability to the *in vivo* human brain. Because of the high resolution and rich (q-t) space coverage, *ex vivo* dMRI of the human brain has found a number of neuroanatomical applications [57] including validation of orientation estimates and tractography [58–61], mapping of subcortical structures and creation of high resolution atlases and templates [4,62,63], and investigation of deep or cortical gray matter laminar structures [43,64–66]. Challenges that are specific to human brain samples include limitations to medium-to-large bore systems (that often do not have specialized hardware such as strong gradient sets), and the need for specific sample holders or coils, although small sub-sections of human brains have also been scanned. Additionally, high quality scans are only possible on well-preserved samples, whereby minimizing the post-mortem interval between death and fixation is a necessity (see Section 4.3 *Fixation* for detailed discussion of the effects of post-mortem interval on MR-relevant tissue features).

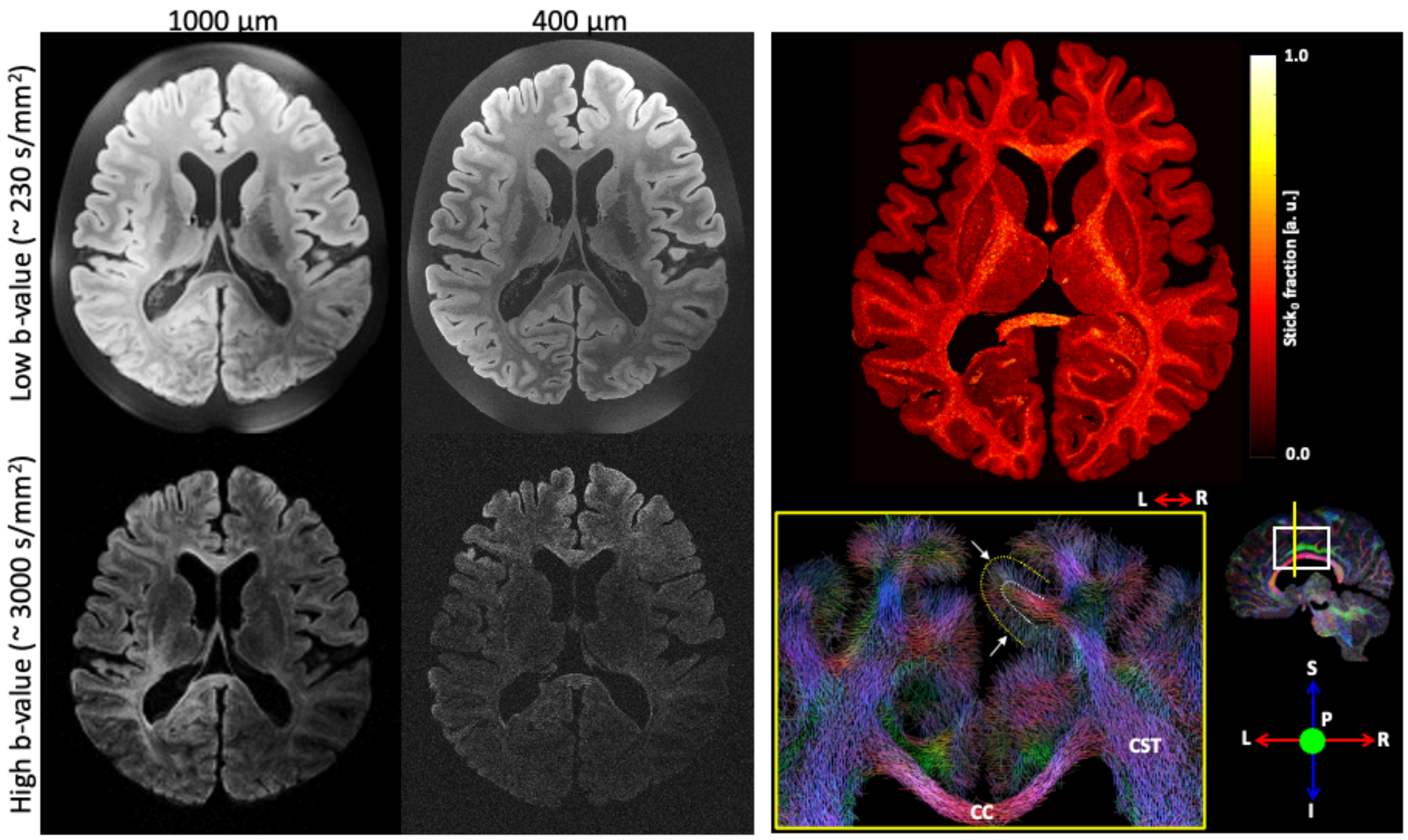

**Figure 5**. *Ex vivo* imaging of the human brain facilitates high resolution and high SNR dMRI (left), which offers exceptional tractography, mapping and creation of templates for small structures, and investigation of gray matter laminar structures (right). Images adapted from [67].

## 3.1.4 Other models

Several other models beyond murine and NHP have proven useful to the diffusion community. Examples include the pig brain, which has been used with *ex vivo* dMRI for tractography validation [68–70], and even to optimize strategies for *ex vivo* diffusion acquisitions [43]. Other gyrencephalic brains have been used *ex vivo* to study diseases or to validate tractography include ferrets [71–73], sheep [74–76], dogs [77], and cats [78]. One study scanned 123 mammalian species, with brains ranging in size from 0.1-1000mL, to study the evolution of mammalian brain connectivity [79].

Further, various non-human models of white matter, including cat optic nerves, rabbit peripheral nerves, garfish olfactory nerves, lobster nerves, and spinal cords were used to evaluate sources of anisotropy, and attribute and associate microstructural features to not only anisotropy, but restricted and hindered diffusion, compartmental diffusivities, and insight into pathology [80,81]. Moreover, the simple nervous system with very large neurons of Aplysia, a sea slug, has been used to study the relationship between the cellular structure and the diffusion MRI signal, to characterize compartment-specific diffusion properties and to follow diffusion changes induced by neuronal responses to ischemic-like stress or chemical stimulation [82–86].

In general, dMRI studies of perfused viable tissue or cells present different advantages and/or challenges as compared to chemically fixed tissue. As mentioned above, isolated perfused 'live' tissue samples facilitate well controlled perturbation studies [86,87], for example controllably changing perfusate tonicity to induce cell swelling/shrinkage to model tissue alterations in stroke[88,89]. Furthermore, the estimation of relative compartment sizes can also be thoroughly explored by controlling tonicity [90], especially since, without fixative, the tissue does not shrink.

# 4 Acquisition

## 4.1 Standard Protocol - overview

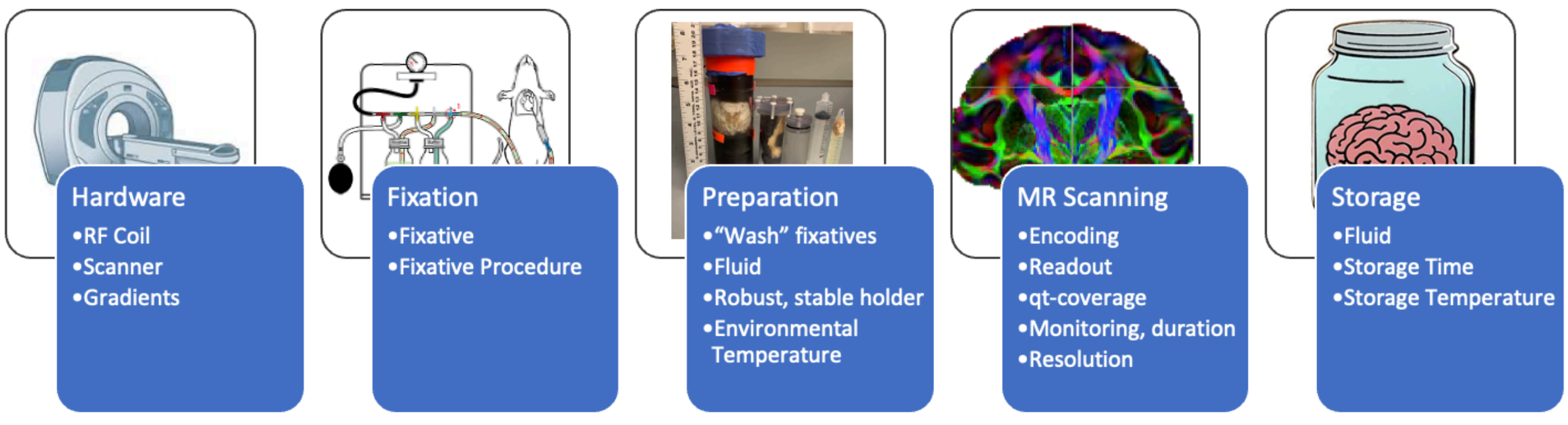

**Figure 6**. For high-quality *ex vivo* diffusion MRI, decisions regarding hardware, fixation, preparation, MR scanning, and tissue storage must be carefully considered. **Hardware**: utilize the smallest coil that fits the sample under investigation, to maximize SNR. **Fixation**: for *ex vivo* tissue to be a good model of *in vivo*, the post-mortem interval to fixation must be as short as possible. **Preparation**: washing out fixative and soaking tissue in a solution of Gadolinium-based contrast agent decreases primarily water-protons $T_1$ which in turn allows for a favorable trade-off between SNR maximization and TR reduction (i.e., reduced acquisition time), while a robust physical setup eliminates motion during scanning. **MR Scanning**: A multi-shot 3D diffusion-weighted spin echo EPI or multi-shot 3D diffusion-weighted RARE/FSE sequence. While not typically used *in vivo* due to motion sensitivity and long scan time, these sequences combine advantages of high SNR, minimal distortion, and reasonable scan time *ex vivo*. **Storage**: fixed tissue can be stored for many months to several years if stored in fixative or phosphate buffered solution (often with 1% PFA) at 5°C.

Here, we provide recommendations for (1) selecting appropriate hardware including MRI systems and coils (Section 4.2), (2) chemical fixation (Section 4.3), (3) sample preparation including washing, constraint and holders, possible contrast enhancements, and immersion solutions (Section 4.4), (4) MR scanning including diffusion encoding, readouts, *q-t* coverage, spatial resolution, and monitoring (Section 4.5), and (5) long term sample storage (Section 4.6).

## 4.2 Hardware (species/organ specific)

Most investigators are limited to the use of hardware available at their imaging center. The guidelines for *ex vivo* imaging largely follow those suggested for small animal imaging in Part 1 [91]. **For RF Coils, the general recommendation is to utilize a coil that will maximize SNR for the sample of interest**, which will typically be the smallest coil that fits the sample under investigation. For *ex vivo* in particular, this is often a volume coil for both excitation and reception, as we typically desire (and have time for) covering the entire sample with the scan field of view, and volume coils have more homogeneous excitation/reception profiles than surface coils. Nevertheless, surface coils for reception can offer higher local SNR so the choice of coil setup depends on the imaging task.

If several MRI systems are available, a recommendation is to **select the scanner with the strongest and fastest gradients, highest field strength, that has a bore and appropriate coil that is large enough for the sample to be imaged**. Higher static magnetic field strengths provide higher SNR, but are challenged by changes in relaxation rates, for example increases in $T_1$ [92] and decreased $T_2$ [93] (see *Sample Preparation* for discussion on techniques to alter both longitudinal and transverse relaxation, and see *Diffusion sequences*: Readout for discussion on taking advantage of altered relaxation rates).

## 4.3 Fixation

After death, tissues begin a self-degeneration process called autolysis, due to their own autogenous enzymes. This process degrades tissue quality, potentially altering several of the microstructural features we wish to quantify and/or validate against. Chemical fixation stops autolytic processes and preserves tissue structure by cross-linking proteins [94]. The postmortem interval (PMI) - i.e. the time between death and chemical fixation - is crucial for tissue quality. Anatomical and radiological signs of autolysis, such as myelin loosening, decreased anisotropy and decreased diffusivities, may be observed as rapidly as four to six hours postmortem, dependent upon tissue temperature, and continue to be altered with longer times between death and fixation [35,94,95]. Thus, rapid tissue fixation is recommended to maintain its integrity.

For laboratory animals, the method of choice is intracardiac **perfusion fixation**, which consists of using the intact vascular system to flush fixatives throughout the tissue upon animal sacrifice. This comes with the advantage of a mostly homogenized fixative distribution throughout the tissues and an efficient removal of blood by flushing with heparin-containing phosphate buffered saline (PBS) prior to perfusion with fixative.

For human tissues or in the case where perfusion-fixation is not possible for animals, perfusion fixation can sometimes be performed on post-mortem brain tissue using mechanical means [57,96].

Alternatively, tissue may be **immersion-fixed**, i.e. immersed in a fixative solution where fixative will passively diffuse throughout the tissue over a period of time (estimated at ~1 mm/hr at 25°C [94], or longer for refrigerated samples). For example, at least 20 days are needed for enough formaldehyde to diffuse to the core of a human brain to cause fixation (Dawe et al. 2009). Other tissue types may require different fixation durations, since the diffusion dynamics of the fixative solution are not the same as for the brain (e.g., 124 days for 4% formaldehyde to penetrate 30 mm into the whole human spleen, and 62 days for the fixative to diffuse over the same distance into mammalian liver tissue) (Dawe et al. 2009). As a major pitfall, immersion fixation introduces the risk of autolytic effects and microbial degradation [97] due to the PMI and/or delay in penetration of deep tissue in larger samples [98]. As autolytic processes are temperature dependent, refrigerating the tissue as soon as possible after death, and during immersion fixation is recommended. Furthermore, this technique yields a transient concentration gradient of the fixative, and hence spatially inhomogeneous tissue integrity with varying MR characteristics (e.g., apparent $T_1$ and $T_2$). As a result, immersion fixed samples may exhibit lower anisotropy and diffusivity than their perfusion-fixed counterparts, in which case the differences may correlate with tissue degeneration. A rule of thumb to check the quality of fixed tissue (perfusion and immersion fixed) is that FA values should be the same as in vivo, e.g. as measured in the midsagittal corpus callosum. Nonetheless, immersion fixation holds the advantage of not relying on an intact vascular system, such that the most peripheral parts of the tissue can undergo good fixation even in the case of clogged blood vessels and/or trauma. Furthermore, perfusion-fixed tissues may additionally be post-fixed through the immersion process as well, often in a different concentration of the fixative.

**In brief, we recommend minimizing the PMI to minimize autolytic changes in the tissue, refrigerating samples before and during immersion fixation and, whenever possible, sacrificing the animals using perfusion-fixation, which practically reduces the PMI to zero**. For samples where perfusion fixation of the tissue is not possible (e.g. human brain samples) the recommendation is to keep PMI as constant as possible across samples and to state the PMI when reporting methods and results.

There are a number of fixatives to choose from, the most common being formaldehyde. For an overview of formaldehyde fixatives, see [99,100]. Notably, the type and concentration of fixative can have a considerable impact on tissue relaxation properties [35,95], leading to differences in image SNR for dMRI. Neutral buffered formalin (NBF; formalin buffered with PBS) at 10% concentration can be used at room temperature and is most commonly used for immersion fixation in large samples (where higher temperature speeds up the penetration), while PFA at 4% in buffered solution should be kept cold and is standard for perfusion fixation. Both result in a 4% formaldehyde solution. Buffered fixatives (NBF or buffered PFA) may be preferred to their unbuffered versions. For example, brains fixed with NBF have higher $T_2$ values than brains fixed with standard formalin ([101] - Figure 5). Reducing these fixative concentrations by half has been shown to also prolong $T_2$ and improve SNR in fixed tissue [102,103]. The addition of glutaraldehyde to the fixative solution improves ultrastructural brain tissue preservation in the case of immersion-fixation[104], a mandatory requirement particularly for electron microscopy cross-validation studies. The use of a fixative solution combining glutaraldehyde and paraformaldehyde has been shown to be MRI compatible while providing a better preservation of the cytoskeletal structures than paraformaldehyde fixative alone [105], while also better preserving membrane permeability [35]. However, high glutaraldehyde concentrations can reduce the immunogenicity of antigens for immunohistochemistry analyses [100], but concentrations in the 0.05% range are acceptable[106]. For an extensive discussion on other types of fixatives, as well as a comprehensive review on fixation in brain banking, see [97].

Even using the recommended procedures and concentrations for perfusion-fixation, there is discrepancy in the literature as to the degree of microstructural changes that the tissue undergoes. Studies mention for example variable levels of **preferential shrinkage** of certain compartments (e.g. the extracellular space) [107]. Brains stored in formaldehyde-based fixatives may continually shrink during storage [108], with different structures experiencing differing rates of morphometric change and with different extents over time [108,109]. It should be noted though that shrinkage due to chemical fixation is less than that during dehydration and tissue preparation for electron microscopy, for example [32,94,107]. At the intracellular level, MR microscopy studies on immersion-fixed, isolated neurons from *Aplysia californica* show that formaldehyde affects nucleus and cytoplasm evenly [83]. Notably, compartment models of diffusion *in vivo* have long yielded relative intra- to extracellular fractions of 30/70 (or 50/50 at best in white matter) which are in mismatch with 80/20 histological estimates of intra- vs extracellular compartment volume fractions [110,111]. The latter are however more consistent with *ex vivo* diffusion models, which typically report 70/30 signal fractions [112]. This suggests that *ex vivo* fixed tissue used for dMRI is closer to its histological counterpart than *in vivo* tissue. It is unclear though whether the change in relative dMRI compartment sizes between *in vivo* and *ex vivo* is due to non-uniform shrinkage with fixation or to non-uniform changes in compartment $T_2$'s, which affect the weighting of compartment signal contributions [113]. Methods such as cryo-fixation used for electron microscopy, which preserves the *in vivo* tissue compartment sizes more faithfully, could help shed light on some of these open questions [107,114].

Furthermore, there is still controversy as to whether chemical fixation increases or decreases **cell membrane permeability** [38,115,116], which is highly relevant for multi-compartment tissue models [112,117]. A number of dMRI studies on fixed tissue have also reported an additional signal component in tissues, such as "isotropically-restricted water" in white matter (sometimes referred to as "still water" or "dot compartment") that is not observed *in vivo* [1,118,119] except for the cerebellum[120]. This isotropically-restricted component is characterized by an extremely low diffusion coefficient, which gives rise to a non-vanishing diffusion-weighted signal even at high *b*-values (Alexander et al 2010). The exact origin of such a component is unknown, although it may be related to tissue overfixation or to vacuoles as visualized using synchrotron imaging[121].

In addition, fixation alters the relaxation rates, substantially decreasing $T_1$ and $T_2$ [35]. The decrease in $T_1$ is understood to be due to the cross-linking of proteins that occurs in the fixation reaction, and is not reversible, whereas $T_2$ is decreased due to the presence of unbound fixative, and can be increased back closer to its *in vivo* value by washing[35]. Consequently, tissue washing or rehydration is beneficial for SNR enhancement and must be considered when designing any *ex vivo* MRI acquisition protocol (see Section 4.4).

Finally, and crucially, fixation changes the water diffusion coefficient[35,95,122,123], with important implications for acquisition protocols and biophysical models of dMRI, as will be discussed in Section 4.5.3.

## 4.4 Sample Preparation

Recommendations for sample preparation include (1) wash out free fixative, (2) ensure a robust mechanical setup to eliminate motion during scanning, and (3) verify temperature stabilization prior to acquiring data. If no tissue relaxometry or diffusion quantification is planned, (4) soaking in a Gadolinium-based solution might help to optimize the trade-off between SNR and acquisition time[39,103]. We discuss each in detail below, and also consider the solution that samples may be scanned in.

First, prior to imaging fixed tissues that have been stored in the fixation fluid, **we recommend PBS 'washes' to rehydrate the sample and wash out free fixatives**. This simply entails placing the sample in a PBS solution, augmented with an antibacterial/antifungal product such as sodium azide, and replacing this PBS solution regularly. While the $T_2$ rises quickly, the washing time required for it to stabilize is dependent on sample size, geometry, volume and temperature of PBS solution, fixative concentration and previous time in fixative[103], with a wide variety of wash times noted in the literature (**Table 1**). To maximize $T_2$ and reduce the time to its stabilization, we recommend using as large a volume of PBS solution relative to the sample volume as practical, and replacing the solution more frequently in the first few hours/days/weeks of washing (e.g., every 12 — 72 hours), depending on sample size. To maximize tissue quality, we recommend refrigerating samples throughout the washing period. Of course, the larger human brain may require several weeks or more to fully wash.

| Specimen | Approximate sample thickness | Soaking time | Fixative | PBS vol | Temperature | Number of solution changes | Reference |
|---|---|---|---|---|---|---|---|
| **Ghost erythrocytes** | 80 µl | 12 hours | various | 100 x sample vol | | 3 | Thelwall et al., 2006, MRM |
| **Rat cortical slices** | 0.5 mm | 12 hours | 4% formaldehyde | | Room temperature | 4-5 | Shepherd et al., MRM, 2009; Shepherd et al., NI, 2009 |
| **Rat spinal cord** | ~2 mm | Overnight | 4% formaldehyde | | | | Shepherd et al., NI, 2009 |
| **Marmoset brain sections** | 2.5 mm | 4 days | 10% formaldehyde | 10 x sample vol | 4°C | 0 | D'Arceuil et al., 2007, NI |
| **Marmoset brain** | ~ 20 mm | 4-6 weeks | 4% formaldehyde | | 4°C | 0 | Blezer et al., 2007, NMR Biomed |
| **Rat brain in situ** | ~ 25 mm | ≥ 20 days | 2% formaldehyde | 50 mL | 4°C | 0 | Barrett et al,. 2022 |
| | | ≥ 47 days | 4% formaldehyde | 50 mL | 4°C | 0 | Barrett et al,. 2022 |
| **Macaque brain** | ~ 40 mm | ≥ 25 days | 10% formaldehyde | 1 L | 4°C | 0 | D'Arceuil et al., 2007, NI |

| | | 3 weeks | 4% formaldehyde | | | | Schilling et al., 2018, NI |
|---|---|---|---|---|---|---|---|
| **Sheep brain** | ~ 50 mm | ≥ 3 weeks | 4% formaldehyde | | | | Leprince et al., Proc ISMRM, 2015 |

**Table 1**. Examples of soaking times in PBS prior to MR imaging, depending on specimen size, fixative, etc.

**To further increase SNR and CNR, gadolinium-based contrast agents are often added during the rehydration (washing) step**. Gd is a paramagnetic contrast agent in the form of a chelate that facilitates longitudinal relaxation (reduces $T_1$), which allows TR minimization, particularly in 3D acquisition sequences, thus maximizing the SNR per unit time. Alternatively, Gd can be introduced during the perfusion step for small animals, in a technique referred to as 'active staining' [124]. If this is done, we still recommend immersing the sample in a Gd solution in addition. Typically, $T_1$ decreases are observed in the sample within 2-3 days in larger brains (macaque brain)[39], although we recommend soaking for 1-2 weeks in Gd solution depending on brain size. Naturally, a higher concentration of Gd results in shorter $T_1$ and $T_2$ [39,103]. The optimal concentration to use depends on acquisition timings (minimum TE, maximum TR) and on the method of staining (active staining typically uses a higher concentration than soaking), as well as field strength. Selected examples from the literature highlight the range of Gd contrasts and concentrations utilized with different acquisition choices given in **Table 2**.

| Concentration (mM) | Contrast agent | TE (ms) | TR (ms) | Magnet (T) | Sample | Reference |
|---|---|---|---|---|---|---|
| 1 | Gd-DTPA | 32 | 240 | 4.7 | Macaque | D'Arceuil et al., 2007 [39] |
| 1 | Gd-DTPA | 41 | 410 | 9.4 | Squirrel Monkey | Schilling et al., 2019 [125] |
| 0.5/0.5 | Gd-DTPA | 26 | 1000 | 9.4 | Zebra Finch | Hamaide et al., 2016 [126] |
| 5 | Gadoteridol | 11-15 | 125-150 | 11.7 | Mouse | Tyszka et al., 2006; Tyszka & Frank, 2009 [127] |
| 5/2.5 | Gd-DTPA | 21 | 100 | 7 | Rat | Johnson et al., 2012 [128] |
| 15/1 | Gd-DTPA | 27 | 250 | 9.4 | Rat | Barrett et al., 2022 [103] |
| 50/5 | Gd-DTPA | 15 | 100 | 9.4 | Mouse | Calabrese et al., 2015 [129] |

**Table 2**. Examples of the concentration of gadolinium contrast agent used in ex vivo dMRI studies from the literature. Where two concentration values are given, the first refers to the active staining concentration (contrast agent added in perfusion fixation), the second to passive staining (soaking post-fixation). A single concentration value refers to passive staining only.

As a caveat, the effects of gadolinium on specific tissue compartments are not well understood. $T_1$ longitudinal evaluation after Gd-doping exhibited a complex behavior of $T_1$ variation within different regions [130]. Gd-soaking is thus best suited for tractography acquisitions, rather than quantitative dMRI studies. Gd staining to optimize SNR has been used without detrimental effects on histology or immunohistochemistry analysis [103,131].

Our second recommendation is to **ensure a robust mechanical setup to eliminate motion** during MRI data acquisition (**Figure 7**). The sample must be tightly constrained inside the imaging container, to prevent motion within the immersion fluid. This is critical to address both bulk motion and non-linear deformations that may arise over the duration of a long *ex vivo* scan (for example, bending of the brain stem). Notably, non-linear deformations are particularly challenging to correct when considering the importance of directional information in diffusion MRI. Sample-holders may be as simple as test tubes, or the cylinder of syringes (which may facilitate removal of air bubbles), as well as holders custom-made to fit within specific volume coils, often made with clear polycarbonate materials. In addition, some fluids that the sample is immersed in (see below) may have high densities, causing the sample to float if not properly constrained. Constraints are often applied through inserting foam pieces, where several groups have suggested very heavily reticulated foam to facilitate removal of air bubbles. Pieces of agar can also be used to stabilize tissue in a container, though agar is MR visible. If the above options are not feasible, post mortem brains can also be placed in a plastic bag. Here, the brain is wrapped with a thin layer of gaze in minimal fluid to reduce air bubbles and susceptibility effects, and the bag is slowly compressed until fluid is expelled to fully remove air bubbles, and tied tightly to ensure no leakage. Finally, very fragile tissue specimens, such as embryonic mouse brains, can be immersed in either agar gels, or kept within the skull, for increased mechanical support and to avoid deformation.

To minimize field inhomogeneities, ideally the holder will consist of a simple geometric shape (without sharp boundaries), with the sample immersed in a susceptibility matched fluid . The holder can be designed to minimize coil-to-sample distance, and to enable consistent sample positioning to eliminate any orientation effects and facilitate analysis.

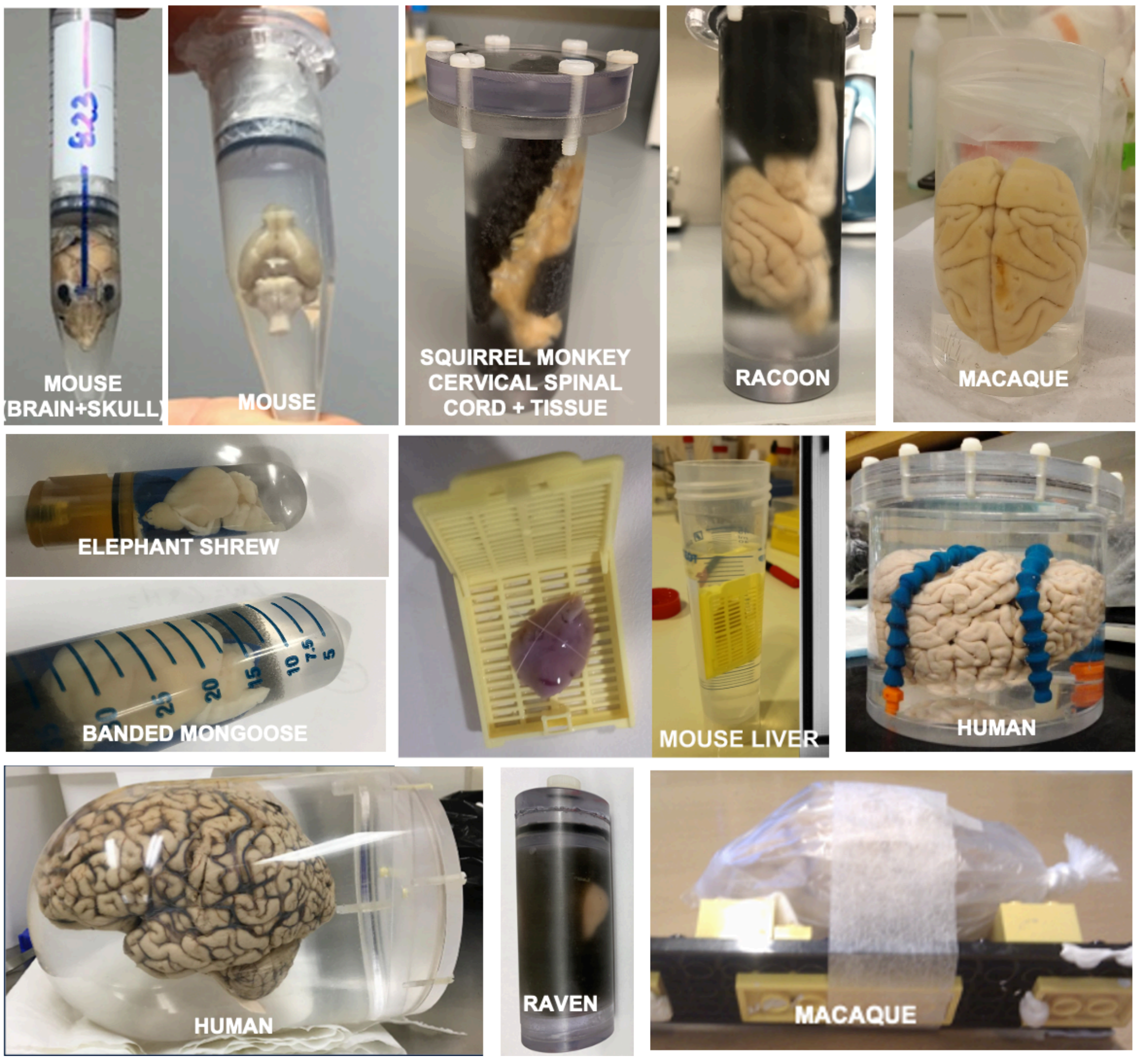

**Figure 7**. Examples of *ex vivo* samples prepared for dMRI acquisitions. Sample holders may be syringes or test tubes/falcon tubes, custom-made or 3D printed holders with ventilation valves, or simply placement within a plastic bag robustly secured to a platform. Photos courtesy of Daniel Colvin, Kurt Schilling, Luisa Ciobanu, Stijn Michielse, Francesco Grussu, Raquel Perez-Lopez, Ileana Jelescu, Tim Dyrby.

Finally, recent advances allow 3D printing dedicated sample holders for minimizing motion/vibration, larger holders for human brains, and the ability to load multiple samples for simultaneous imaging (although with the disadvantage of requiring larger FOV, limited spatial resolution and sub-optimal shimming across samples).

For data acquisition, samples are typically immersed in either PBS or fluorinated oil. **In general, we suggest using a fluorinated oil, which are susceptibility-matched, inert compounds that do not have $^{1}$H protons**. Thus, they lead to no signal in $^{1}$H MR images, alleviating ghosting artifacts, facilitating image masking, and allowing a smaller, tighter FOV. A variety of fluorinated oils are suitable: perfluoropolyethers such as Fomblin (Solvay) or Galden (Solvay) or perfluorocarbons such as Fluorinert (3M), and can be recycled to be reused for multiple scans. While studies have not explicitly looked at the effects of these oils on conventional or immuno-histology, it is our experience that they do not compromise nor interfere with this analysis [132]. However, it is recommended that after the MRI scan, excess oil be removed off the sample using absorbent paper, followed by several PBS washes to remove any residual oil in the cavities of the tissue sample. While these compounds are inert, the effect of long term storage of tissue in them is not known and thus not recommended.

Due to the strong difference in magnetic susceptibility with tissue, any air bubbles trapped in cavities such as brain sulci and ventricles will result in substantial image distortions and should be carefully removed during sample preparation by slightly turning, shaking and agitating the tissue. Alternatively, a vacuum setup to remove trapped air bubbles is very effective. Of note, due to the extremely low pressure, bubbles expand substantially, and the fluid can appear as 'boiling'. Lastly, a fine paintbrush may also be used to remove air bubbles while the sample is immersed.

Finally, to remove unwanted time-dependent signal contributions due to tissue temperature changing during scanning (from the cold storage to bore-temperature), **we advise placing the sample at the desired temperature for at least 4-8 hours prior to scanning** [43]. It may be beneficial to run a dummy dMRI scan during these hours of temperature regularization to also stabilize possible temperature changes caused by gradient coil heating. As nonlinear motion of the tissue (due to physically handling and setting up the tissue in the magnet) is likely to occur immediately after handling, this extra waiting period (or dummy scans) may minimize motion during the scan itself. Timing should be kept consistent to avoid bias in group comparisons. During the scanning, the use of strong or varying diffusion weightings can influence the temperature environment of the tissue hence the diffusion coefficient. A constant airflow around the tissue can further stabilize the temperature.

## 4.5 MR Scanning

### 4.5.1 Encoding

Just as for *in vivo* small animal imaging (Part 1 [91]), a number of possible **diffusion encoding**, or sensitization schemes are feasible, although the unique changes of decreased diffusivities and relaxation times must be considered. For *ex vivo* dMRI, the two most common encoding schemes are the **pulsed gradient spin echo (PGSE)**[133] and **steady-state free-precession (SSFP)** with diffusion preparation[134]. For **PGSE** encoding, strong diffusion-sensitization gradients are applied on either side of a 180° refocusing pulse, resulting in a mathematically elegant way to describe diffusion weighting through the *b*-value, $b = (\gamma G\delta)^2(\Delta - \delta/3)$, (where $\gamma$ is the gyromagnetic ratio, *G* is the gradient strength, $\delta$ is the pulse duration, and $\Delta$ pulse separation). Because of its simplicity, PGSE is the most widespread diffusion weighting in both *in vivo* and *ex vivo* experiments.

Additionally, SSFP has been heavily utilized *ex vivo*. In **SSFP** acquisitions, the signal is retained over multiple repetition times, generating both spin echoes and stimulated echoes simultaneously. This results in both high SNR in the absence of diffusion gradients and in strong diffusion weighting in their presence. The primary advantage of SSFP is its highest SNR per unit time of all sequences, and its primary disadvantage is increased sensitivity to motion, limiting its use *in vivo*, but easily overcome *ex vivo* [134]. However, unlike PGSE, the signal becomes a complicated function of flip angle, TR, $T_1$, $T_2$, and the diffusion encoding, requiring specialized modeling to quantify diffusion coefficients[135].

Other encodings are also possible *ex vivo*, and described in more detail in Part 1. Briefly, **stimulated echo acquisition mode (STEAM)** sequences enable probing very long diffusion times with the disadvantage of half the SNR compared to PGSE (for equal echo times)[67]. **Oscillating gradient spin echoes (OGSE)** uses periodic sinusoidal gradients to probe much shorter times and length scales, which become particularly small with reduced diffusivity *ex vivo*, with limitations associated with attaining higher *b*-values[136,137]. Finally, diffusion encoding can be applied along multiple spatial directions in **multi-dimensional diffusion encoding** experiments [138], which offers potential contrasts related to compartmental kurtosis, compartmental exchange, microscopic anisotropy, or heterogeneity of structural sizes/diameters.

## 4.5.2 Readout

For ex vivo imaging, readouts can be extremely diverse due to lack of sample motion and long available scan time. Our recommended starting protocol is a **multi-shot (segmented) 3D EPI sequence**. While this acquisition scheme comes at the cost of increased scan time and possible artifacts due to physiological motion in vivo vs. its 2D single shot counterpart, conveniently, ex vivo MRI is not limited by scan time nor motion, hence the recommendation. For this reason, it is not uncommon to use 4-12 segments, or more. The number of segments should ideally be chosen with the matrix size in mind so that segments of equal size are acquired.

Here, two features have been changed from the typical *in vivo* protocol: from single to multi-shot, and from 2D multi-slice to a 3D sequence. 3D EPI sequences are capable of achieving substantially higher SNR than 2D EPI [41,139]. This SNR gain is due to averaging effects from Fourier encoding the entire tissue volume (i.e., collecting signal from the entire 3D tissue volume). SNR increases as the square root of the number of datapoints in the third spatial dimension, also referred to as second phase-encode direction, and which corresponds to the slice dimension in 2D multi-slice experiments[140]. The rationale for going from single to multi-shot is that a strong segmentation of the 3D EPI read-out is necessary to prevent image distortions and prohibitively long echo times, especially if high spatial resolution is desired. 3D EPI also comes with the advantage of enabling acquisitions with truly isotropic resolution. Indeed, very high spatial resolutions in the third dimension may not be achievable in 2D multi-slice acquisitions, as very thin slices may not be feasible depending on the slice-selection gradient strength.

Despite the intrinsically higher SNR of 3D images, they suffer from suboptimal temporal utilization of $T_1$ relaxation. The TR with optimal SNR per unit time depends on the $T_1$ value of the tissue and can be determined by means of Bloch simulations or using the following relationship: $SNR_t \sim M_0\left(1 - e^{-TR/T_1}\right)e^{-TE/T_2}$. This equation can be used to determine the optimal TR in terms of **SNR efficiency** (SNR / √(total imaging time)): $SNR_{efficiency} \sim (1/\sqrt{(TR)})\left(1 - e^{-TR/T_1}\right)$. Conveniently, *ex vivo* imaging enables the addition of Gadolinium-based contrast agents to the sample to shorten $T_1$, and thus the optimal TR, facilitating high SNR efficiency. Example SNR efficiency curves are shown in **Figure 8**, where the optimal TR is ~1.25 times the estimated $T_1$ (for examples of optimizing TE, TR, and diffusion weighting see *Other Considerations* in Section 4.5.3 *q-t coverage*; for a discussion of advantages and limitations of contrast agents, see *Sample Preparation Section 4.4*). It should be noted however that short TR's may introduce $T_1$ weighting and affect the relative contributions of different compartments to the overall signal, in a similar manner to $T_2$ weighting.

Alternative readouts are also possible *ex vivo*. This includes the **2D EPI** and **spiral** readouts as *in vivo* (described in detail in Part 1 [91]), and many *ex vivo* studies have taken advantage of the multiple RF echo trains of **RARE/FSE**, or **GRASE**, which often offers a good trade-off between scan time and image quality with high SNR and immunity to distortions. At the extreme end of acquisition is a **line-scan** readout, which traverses a single line of *k*-space per excitation. While this offers excellent robustness to susceptibility artifacts, it has a low SNR efficiency and may require excessively long scan times, or a tradeoff between spatial and angular resolutions. $B_0$ drift may also become problematic when the scan time is on the order of several days, and should be corrected for. Acquisitions can for example be split into 2-3 hour blocks, with the frequency (and shim, potentially requiring a new B0 map to be acquired) readjusted before each block. This also allows retrieving at least partial data in the event the scanner crashes during the long acquisition.

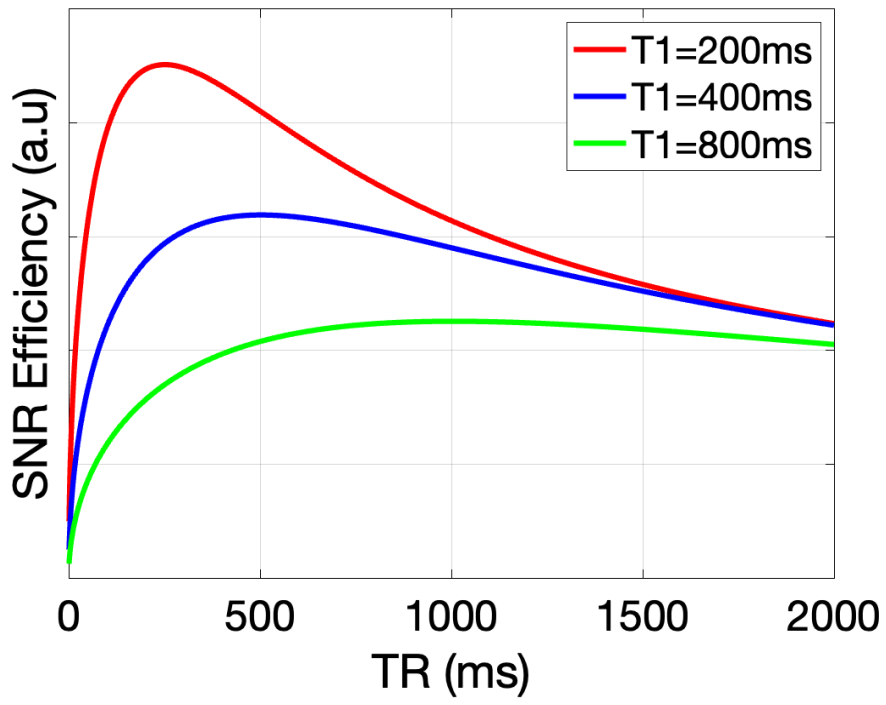

**Figure 8**. Plots showing how SNR efficiency varies with T1. Curves are based on the SNR-efficiency equation given in Section 4.5.2 based on a spin echo sequence. The optimal TR is ~1.25 times the sample T1, although there is a wide range of near-maximum efficiency. Similar optimization can be performed for TE, and diffusion weightings (see *Other Considerations* in *Section 4.5.3 q-t coverage* for examples)

## 4.5.3 q-t coverage

Before setting up diffusion parameters, it is important to understand the theoretical requirements of the chosen data analysis framework which will be used to process the data - this might include requirements of short gradient pulses, long/short diffusion times, certain b-value regimes, or a number of unique sampling directions.

The primary differences between *in vivo* (described in Part 1) and *ex vivo* data acquisition is the slower diffusion in fixed tissue, which often requires high b-value and/or longer diffusion times to ensure adequate signal attenuation and spin displacement: concretely, *b*-values should be adapted to ensure ~constant *bD* product[39,57] (we assume here that the sample has been washed to restore $T_2$ values, otherwise short $T_2$ is another consideration for fixed tissue). Indeed, *ex vivo* diffusivities of fixed tissue remain lower than *in vivo* even at 37°C, the degree depending on the fixation method (perfusion-fixation or immersion) and postmortem interval [35,39,43,57,141,142]. Moreover, scanning is often performed at room temperature, resulting in further reduction of diffusivity (change of -1–3% per °C). Selected examples are given in **Table 3**, where considerable decreases in diffusivity are observed across species in perfusion and immersion fixed brains, as well as in fresh ex vivo tissue. Notably, for tissue that has been suitably perfusion-fixed, the impact of fixation on the diffusivity is relatively comparable across samples/subjects.

| | **T1** | **T2** | **D** | **Specimen** | **Reference** |
|---|---|---|---|---|---|
| **Fixation (% change from in vivo to fixed)** | | | ↓ 80% ADC | macaque brain 2.5mm sections, immersion fixed | D'Arceuil et al., 2007, NI |
| | | | ↓ 64% Trace | mouse brain perfusion fixed | Sun et al, 2003 MRM |
| | | | ↓ 55% MD | Rat spinal cord perfused fixed | Madi et al, 2005, MRM |
| | | | ↓ 72% Trace | mouse brain perfusion fixed | Sun et al, 2005, MRM |
| | | | ↓ 50% MD | squirrel monkey, perfusion fixed | Schilling et al, 2017, MRI |
| | | | ↓ 62% MD | rat brain, perfusion fixed | Wang et al., Eur Radiol Exp, 2018 |
| | ↓ 22% | ↓ 5% | ↓ 48% MD | Marmoset brain, immersion | Haga et al., Magn Res Med Sci, 2019 |
| | ↓ 40% | ↓ 3% | | mouse brain perfusion fixed | Guilfoyle et al, 2003, NMR Biomed |
| | ↓ 63% | ↓ 35% | | human immersion fixed | Pfefferbaum et al, 2004, NI |
| | ↓ 69% | ↓ 27% | | human immersion fixed | Birkl et al, 2016, NMR Biomed |
| **Ex vivo fresh (% change from in vivo to ex vivo)** | | | ↓ 65-88% MD, AD, RD | fresh pig brain | Walker et al, 2019, PLoS One |
| | | | ↓ 50% ADC | fresh monkey brain | D'Arceuil et al., 2007, NI |
| **PBS washing (% change from fixed to washed)** | ↑ 7% | ↑ 30% | ↑ 30% ADC | macaque brain 2.5mm sections, immersion fixed | D'Arceuil et al., 2007, NI |
| | ↑ 7% | ↑ 3% | ↑ 11% MD | human immersion fixed | Leprince et al, 2015, Proc ISMRM |
| | ↑ 3% | ↑ 72% | ↑ 2% MD | rat brain, perfusion fixed | Barrett et al n.d. |
| | | | Reported no difference | Cat spinal cord, immersion fixed | Pattany et al, 1997, AJNR |
| | Reported no difference | ↑ 516% (cortex) | | rat cortical slices, immersion fixed | Shepherd et al., 2009, MRM |
| | ↓ 14% | ↑ 24% | | Marmoset brain | Blezer et al, 2007, NMR Biomed |

**Table 3**. Changes in $T_1$, $T_2$ and $D$ reported in the literature due to fixation and washing in PBS. Data is included from samples fixed with 10% Formalin or 4% PFA, scanned at room temperature. Measurements from WM only are included, unless otherwise noted.

Thus, for both fresh *ex vivo* and fixed tissue the drop in diffusivity is typically on the order of ~2-5 at room temperature, which corresponds to increasing the *b*-value by a similar factor to match the attenuation expected from an *in vivo* dMRI protocol. For postmortem human samples, however, due to non-negligible post-mortem interval and an extended duration of immersion fixation needed to preserve cell and tissue components, the diffusivity is often on the order of 85% lower than *in vivo* [143,144], also in agreement with animal studies with extended PMI [95]. Consequently, going beyond DTI protocols, including DKI and other advanced dMRI methods is sometimes challenging on postmortem human tissue due to the prohibitively high *b*-values required. For DTI, an adjusted *b*-value to about 4000 s/mm$^2$ [39] has been shown to result in similar signal attenuation as for an *in vivo* *b*=1000 s/mm$^2$ scan, and provide the angular contrast needed to resolve crossing fibers for tractography [43]. It should, however, be noted that the optimal *b*-value for post-mortem acquisitions is a function of tissue fixation and scanning temperature and should be evaluated for each experiment individually.

Because of changes in diffusivity, it is important to consider spatial scales that are being probed. As diffusivity drops, diffusion distances proportional to $\sqrt{Dt}$ drop as well - unless the diffusion time is prolonged accordingly - which results in different interactions between water molecules and the microscopic features they are able to probe. This may be beneficial when interested in probing geometry on small scales and also extends the limit of the narrow pulse approximation validity[1]. This is additionally important in MR microscopy, where the resolutions start to approach the diffusion length scales and a significant amount of water may diffuse out of the imaging voxels in the echo time such that the spatial resolution is no longer 'real'.

Below, we provide guidelines for common applications of ex vivo imaging: signal representations (DTI/DKI), tractography, and biophysical signals models.

**Diffusion tensor imaging (DTI)**: as in in vivo, recommendations include 20-30 non-collinear directions to mitigate noise, and b-value chosen so that the product $bD \simeq 1$ to maximize precision. Ex vivo, this results in a b-value of approximately 2500 - 5000 s/mm$^2$, depending on the drop in diffusivity values. For Diffusion Kurtosis Imaging (DKI) the radius of convergence of the cumulant expansion ex vivo translates into the highest b-value roughly double the optimal one for DTI[8,12,145,146] with a recommended 20-30 directions per shell.

For **tractography**, ex vivo guidelines again follow closely those of both small animal (Part 1) and human scans in vivo - our recommended protocol includes acquiring 50-60 directions at a moderate-to-high b-value, where a greater diffusion weighting (particularly for ex vivo) leads to a higher angular contrast and ability to resolve complex fiber architectures - for example Dyrby et al.[43] found a b-value of ~4000 s/mm$^2$ to lead to consistent fiber reconstructions with a high angular contrast, although a much larger range of b-values has also been utilized ex vivo with high angular accuracy and subsequent accurate tractography[59,147–150].

Regarding **compartment modeling**, it is critical to consider the data requirements of the intended biophysical model, specifically as it relates to diffusion times and diffusion weightings. Similarly, if diffusion-relaxometry experiments are intended, altered $T_1$ and $T_2$ depending on fixation and subsequent washing must be considered.

**Other practical considerations**

A great benefit in ex vivo imaging is the ability to conduct multiple experiments on the same tissue. Because relaxation and diffusivity can vary dramatically based on sample preparation (rehydration time, fixative, fixative concentration, contrast agent concentration, etc.) we recommend informing acquisition settings by acquiring scans prior to the start of a new study and optimizing using methods described in the literature [39,103]. For example, It is particularly advantageous to measure $T_1$, $T_2$, and diffusivity throughout the sample, and **SNR efficiency** can be optimized and tuned for both 2D and 3D sequences through changes in TE, TR, and diffusion weightings (**Figure 9**).

Other practical suggestions, as described in detail in Part 1 and also apply to *ex vivo* scans are to: (1) use an optimally distributed set of diffusion-encoding directions that cover the full sampling sphere (2) randomize ordering of acquisition of DWI images, especially across *b*-values to reduce or enable correction of temporal biases, reduce duty cycle, and allow analysis on partial acquisitions (3) intersperse several/many *b*=0 images throughout the scan to enable controlling for temporal drifts [44], (4) use the effective *b*-matrix (i.e., the realized *b*-matrix taking into account other sequence gradients — rather than the nominal value entered into the scanner) which can be measured and validated in phantoms [151], and (5) perform high order shimming which may be critical for image quality, particularly at high field strengths due to their increased inhomogeneities.

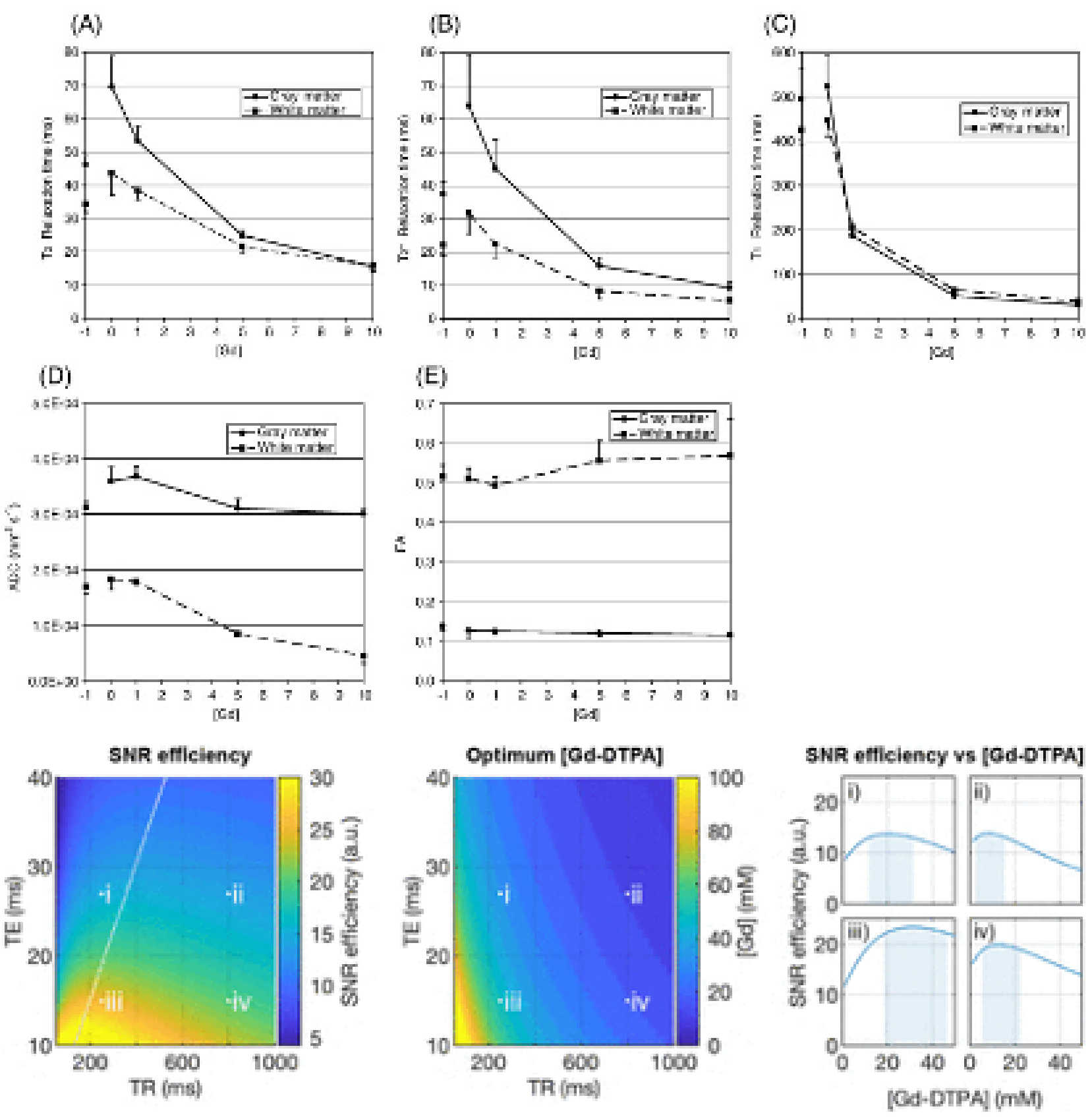

**Figure 9**. Approach to optimizing ex vivo diffusion protocols. Relaxometry (top) and diffusion (middle) can be measured as a function of contrast concentration, or fixative solution, for both white and gray matter tissue types, and SNR efficiency can be optimized (bottom) by manipulating sequence parameters and gadolinium contrast agent concentration. Images are adapted and modified from [39] (top) and [103] (bottom).

## 4.5.4 MR Scanning, Monitoring, scan duration

During long ex vivo scanning, it is important to minimize artifacts using several acquisition strategies. First, it is recommended that each image or volume be collected within as short a time period as possible, i.e., inner-most loops should cover k-space sampling while outer-most loops should cover q-space sampling. Furthermore, if more than a single acquisition is performed using averaging, it is preferred to acquire each image separately split into repetitions, while saving the complex-valued data. Averaging repetitions in complex space rather than magnitude space is beneficial for lowering the noise floor to its theoretical zero-value, as opposed to being affected by the Rician noise floor; however, the advantage of initially saving these volumes as individual repetitions gives the opportunity to omit potentially corrupted individual images, if needed. Phase corrections prior to averaging are mandatory, and may need to be implemented by the user if not addressed well by the scanner reconstruction directly [152,153].

A stable temperature throughout the protocol is warranted to limit undesired signal drift due to $T_1$-weighting and diffusivity variations. Potential sample temperature changes over time should therefore be prevented for example by ensuring an air flow around the tissue with constant temperature, as it is not easy to compensate for temperature drift in the post-processing pipeline. As a temperature control, a water vial can be placed next to the tissue and monitor its diffusivity over time, although it is important to carefully consider where the vial will appear within the field of view due to the addition of possible susceptibility, ringing, or ghosting artifacts.Of note, a temperature increase may also occur *in vivo*, though the effect may be less pronounced thanks to active thermal regulation of the animal.

Collectively, these strategies do not prevent the artifacts, but make them more addressable with subsequent post-processing.

## 4.5.5 Spatial resolution

For in vivo human or small animal imaging, spatial resolution should be as high as permissible for the targeted SNR and available scan time. However, *ex vivo* scans require different considerations when choosing a spatial resolution, due to the substantially increased scan time. While signal drift and temperature stability issues may arise due to increased scan time, resolution can be pushed quite extensively, with recent protocols nearly pushing the boundaries where resolution limits are set by the diffusion process itself, e.g. below 10 micrometers [154].

In short, there is no single set of guidelines, or consensus, on image resolution for specific species, nor for specific experimental designs. Rather than providing specific recommendations for resolution, below we give typical volumes of brains, and compute what the equivalent voxel size (i.e, the **volume equivalent resolution**) would be given the ratio of volumes, and a 2-mm isotropic human brain scan, typical of in vivo studies (**Table 4**). Further, we give examples of *ex vivo* scans that are **pushing the boundaries of resolution** - we note that these are not always feasible at every imaging center, and are not for any specific tractographic or modeling purposes, but only to highlight high resolution scans that have been performed.

### 4.5.5.1 Volume equivalent resolutions and pushing the boundaries

| Species | Brain Volume (mL) | Matching spatial resolution (isotropic) | Reported in literature (*ex vivo*) |
|---|---|---|---|
| Human | 1200 | 2 mm | 730-μm isotropic (b=3000 s/mm$^2$, 64 directions, 4 days, 3D-EPI) [141]<br>940-μm isotropic (*b*~4500 s/mm$^2$, 54 directions, 1m15s per volume, repeated 10 times, DW-SSFP) [144]<br>1000-μm isotropic (*b*~5175-8550 s/mm$^2$, 49-52 directions, DW-SSFP, 10-11 minutes per volume) [143]<br>500-μm isotropic (b~1700-5000 s/mm$^2$, 120 directions, DW-SSFP, 45 minutes per volume) [155]<br>100-μm isotropic (hippocampus) (b=4000 s/mm$^2$, 12 directions, 63hr acquisition)[66]<br>200-μm isotropic (brainstem+thalamus) (b=4000 s/mm$^2$, 120 directions, 208hr acquisition)[63]<br>400-μm isotropic (*b*~3000 s/mm$^2$, 2h23m/volume); 500-μm isotropic (*b*~2000/4000 s/mm$^2$, 53m/volume); 1000-μm isotropic (*b*~1000 s/mm$^2$, 23m/volume) - kT-dSTEAM [67] |
| Mouse | 0.4 | 140 μm | 25-μm isotropic (*b*=4000 s/mm$^2$, 61 directions, 95h scan, 3D-SE) [156]<br>43-μm isotropic (*b*=4000 s/mm$^2$, 120 directions, 235h scan, 3D-SE) [129]<br>100-μm isotropic (b=2000/5000 s/mm$^2$, 60 directions, 12h scan, 3D-SE) [157] |
| Rat | 0.6 | 160 μm | 50-μm isotropic (*b*=3000 s/mm$^2$ ,61 directions, 289h scan, 3D-SE) [158]<br>150-μm isotropic (*b*=3000/6000 s/mm$^2$, 30/30 directions, 21h scan, 3D-DW-GRASE) [159]<br>88-μm isotropic (*b*=800 s/mm$^2$, 12 directions, 18h scan, 3D-SE) [27]<br>78-μm isotropic (*b*=1500 s/mm$^2$, 6 directions) [25] (*b*=4000 s/mm$^2$, 30 directions) [25] |
| Squirrel monkey | 35 | 600 μm | 300-μm isotropic (*b*=3000/6000/9000/12000 s/mm$^2$, 100 directions each, 48hr scan, 3D-EPI) [149] |
| Mini-Pig | 64 | 750 μm | 500-μm isotropic (*b*=4009 s/mm$^2$), 61 directions, 28 hrs, 2D-SE) [70] |
| Macaque | 80 | 800 μm | 390x540x520 μm (*b*=1000 s/mm$^2$, 8 directions, 45hr scan, 3D multiple echo SE) [26]<br>500-1000 μm isotropic (*b*=1477 - 9500 s/mm$^2$, 20-180 directions, up to 19 days scan, 2D-SE) [17,160,161]<br>600-μm isotropic (*b*=4000 s/mm$^2$, 60 directions, 2D-SE single echo) [162]<br>200-μm isotropic (*b*=100-10,000 s/mm$^2$, 3-36 directions, 93hr scan, 3D-EPI) [139,162]<br>200-μm isotropic (*b*=500, 1000, 4000,10 000 s/mm$^2$, 8- 16-32-64 directions respectively, 73hr scan, 3D-EPI) [130].<br>300-μm isotropic (b=6000 s/mm$^2$, 101 directions, 3D-EPI) [125]<br>250-μm isotropic (b=4800 s/mm$^2$, 121 directions, 71hr scan, 3D-EPI) [56]<br>600-μm isotropic (b = 4000 s/mm$^2$, 128 gradient directions, DW-SE multi-slice) [163]<br>1000-μm isotropic (b = 4000, 7000 and 10000 s/mm$^2$, 250-1000-1000 directions respectively + spherical tensor encoding, DW-SE multi-slice) [163] |
| Macaque | 35 | 600 μm | 80-μm isotropic (b=2400 s/mm$^2$, 64 directions, 15 days 3D-EPI) [5] |

**Table 4**. Summary of brain volumes of various species, matching spatial resolutions to the typical one for human dMRI, and ranges of spatial resolutions reported in the literature, *ex vivo*. The references provided are not comprehensive.

MR microscopy is defined as MR imaging with a spatial resolution in the micrometer range, which makes it possible to even image individual cells. Indeed, with dedicated setups that allow sufficiently high SNR, MRI with a resolution of a few μm becomes feasible [83,164–167]. By convention, MRI transforms to MR microscopy (MRM) when the voxel side lengths are less than 100 μm [168]. A number of effects and considerations are encountered in MR microcopy that must be considered with respect to hardware, sequences, diffusion dispersion, and data processing. For an introduction to MRM emphasizing practical aspects relevant to high magnetic fields see [169], a review in [170], and details of microscale nuclear magnetic resonance hardware in [171].

### 4.6 Storage

*Ex vivo* offers the advantage that repeated scans of fixed brains can be performed, enabling longer scans, multiple sessions, and optimization of the sequences and contrasts over time. **Here, if tissue storage is necessary, we recommend storing the samples in either a PBS solution with Sodium Azide to inhibit bacterial and fungal growth, or in a weak fixative solution (1% formalin in PBS), at low temperature (4-5°C)**. Long-term storage at room temperature is also possible in 4% formaldehyde, although it is associated with an increase in formic acid and methanol which may have a dehydrating effect, and a slow decline in T1 and T2 relaxation times [172,173]. These effects can be mitigated by regularly refreshing the formaldehyde solution and by sufficient rehydration prior to scanning. Several studies have investigated the effects of storage time on diffusion metrics, and concluded that with appropriate care, tissue can be rescanned over several years with negligible variability in results [43,174]. Once fixed, the tissue quality should be inspected periodically [32,43] to ensure tissue integrity and absence of bacterial or fungal growth.

# 5 Conclusions

In this manuscript, we have provided an overview of best practices for ex vivo diffusion MRI, focusing on experimental design, sample preparation, and MR scanning. These steps are critical to ensure rigorous and reproducible data collection, and pave the way for subsequent data processing. As we move to the next part of this series, we will shift our focus to pre-processing, model-fitting (processing), tractography, and comparisons with microscopy. Together, these recommendations are given to facilitate high quality studies and interpretation of ex vivo dMRI data.

# 6 Acknowledgements and Support

The authors acknowledge financial support from: the National Institutes of Health (K01EB032898, R01AG057991, R01NS125020, R01EB017230, R01EB019980, R01EB031954, R01CA160620, R01NS109090), the National Institute of Biomedical Imaging and Bioengineering (R01EB031765, R56EB031765), the National Institute on Drug Abuse (P30DA048742), the Secretary of Universities and Research (Government of Catalonia) Beatriu de Pinós postdoctoral fellowship (2020 BP 00117), “la Caixa” Foundation Junior Leader fellowship (LCF/BQ/PR22/11920010), the Research Foundation Flanders (FWO: 12M3119N), the Belgian Science Policy Prodex (Grant ISLRA 2009–1062), the µNEURO Research Center of Excellence of the University of Antwerp, the Institutional research chair in Neuroinformatics (Sherbrooke, Canada), the NSERC Discovery Grant, the European Research Council Consolidator grant (101044180), the Canada Research Chair in Quantitative Magnetic Resonance Imaging [950-230815], the Canadian Institute of Health Research [CIHR FDN-143263], the Canada Foundation for Innovation [32454, 34824], the Fonds de Recherche du Québec - Santé [322736], the Natural Sciences and Engineering Research Council of Canada [RGPIN-2019-07244], the Canada First Research Excellence Fund (IVADO and TransMedTech), the Courtois NeuroMod project, the Quebec BioImaging Network [5886, 35450], the Mila - Tech Transfer Funding Program and the Swiss National Science Foundation (Eccellenza Fellowship PCEFP2_194260), the Wellcome Trust (202788/Z/16/A, 203139/Z/16/Z and 203139/A/16/Z).